\documentclass[aps, pra, twocolumn, superscriptaddress, nofootinbib, reprint, floatfix]{revtex4-1}

\usepackage{amsmath,amssymb}  
\usepackage{amsfonts}
\usepackage{graphicx}   
\usepackage{hyperref}   

\usepackage{color}
\begin{document}

\title{Helicity in Superfluids: existence and the classical limit}
\author{Hridesh Kedia}
\affiliation{James Franck Institute and Department of Physics, The University of Chicago, 929 E 57th St, Chicago, IL 60637, USA}
\altaffiliation{Current address: Physics of Living Systems Group, Massachusetts Institute of Technology,
Cambridge, MA 02139, USA}

\author{Dustin Kleckner}
\altaffiliation{Current address: University of California, Merced,
5200 N. Lake Road
Merced, CA 95343, USA}
\affiliation{James Franck Institute and Department of Physics, The University of Chicago, 929 E 57th St, Chicago, IL 60637, USA}

\author{Martin W.~Scheeler}
\affiliation{James Franck Institute and Department of Physics, The University of Chicago, 929 E 57th St, Chicago, IL 60637, USA}

\author{William T.~M.~Irvine}
\email{wtmirvine@uchicago.edu}
\affiliation{James Franck Institute, Enrico Fermi Institute and Department of Physics, The University of Chicago, 929 E 57th St, Chicago, IL 60637, USA}

\begin{abstract}
In addition to mass, energy, and momentum,  classical dissipationless flows conserve helicity, a measure of the topology of the flow. 
Helicity  has far-reaching consequences for classical flows from Newtonian fluids to plasmas. 
Since superfluids flow without dissipation, a fundamental question is whether such a conserved quantity exists for superfluid flows.
We address the existence of a ``superfluid helicity'' using an analytical approach based on the symmetry underlying classical helicity conservation: the particle relabeling symmetry. Furthermore, we use numerical simulations to study whether bundles of superfluid vortices which approximate the structure of a classical vortex, recover the conservation of classical helicity and find dynamics consistent with classical vortices in a viscous fluid.
\end{abstract}

\maketitle
\section{Introduction}
Our understanding of fluid flow is built on fundamental conservation laws such as the conservation of mass, energy, and momentum \cite{landau_fluid_1987}. In particular, these give rise to the Euler equations of dissipationless fluid mechanics which capture many fluid phenomena including vortex dynamics \cite{christodoulou_euler_2007}, 
instabilities \cite{constantin_euler_2007} and play a key role in the study of turbulence \cite{dombre_chaotic_1986,beale_remarks_1984}.

Hidden within the Euler equations for isentropic flows, is a less familiar conservation law \cite{woltjer_theorem_1958,moreau_constantes_1961,moffatt_degree_1969}: conservation of helicity $\mathcal{H}_\textrm{Euler} = \int \textrm{d}^3x\,\mathbf{u}\cdot\pmb\omega\;,\;\pmb\omega=\nabla\times\mathbf{u}$. As a measure of the average linking of vortex lines \cite{moffatt_degree_1969,moreau_constantes_1961}, helicity conservation places a topological constraint on the dynamics of classical inviscid isentropic flows\footnote{From here on, we refer to classical inviscid isentropic flows as Euler flows}. Helicity has further yielded new insights into viscous flows, from vortex reconnection events \cite{scheeler_helicity_2014,kimura_reconnection_2014},  to the study of coherent dynamical structures generated by turbulent flow \cite{levich_role_1983,hussain_coherent_1986,yokoi_statistical_1993}.

Superfluids\footnote{We shall only consider superfluids with a complex scalar order parameter as in ${}^4$He and atomic Bose-Einstein condensates.} display striking similarities with classical fluids in their vortex dynamics \cite{barenghi_quantized_2001,paoletti_velocity_2008} and turbulence statistics \cite{vinen_classical_2000,vinen_quantum_2002,yepez_superfluid_2009}. Since superfluids flow without dissipation, it is natural to ask whether a  conserved quantity analogous to helicity also exists in superfluid flows. Natural candidates for a ``superfluid helicity'' are: (i) the expression for the classical helicity $\mathcal{H}_\textrm{Euler}$ which is not conserved in superfluid flows \cite{scheeler_helicity_2014,clark_di_leoni_helicity_2016}, and (ii) a Seifert-framing based helicity which vanishes identically \cite{akhmetev_borromeanism_1992,scheeler_helicity_2014,hanninen_helicity_2016,salman_helicity_2017}. However, it has been challenging to establish their connection to the fundamental notion of conservation. It has thus remained unclear whether additional conserved quantities akin to helicity and circulation exist in superfluids, and how a ``classical limit'' of superfluid helicity might behave.

In this letter, we use an analytical approach based on the particle relabeling symmetry, which underlies helicity conservation and Kelvin's circulation theorem in classical inviscid fluids, to address the question of a ``superfluid helicity". We find that the conserved quantities associated with the particle relabeling symmetry in superfluids vanish identically, yielding only trivial conservation laws instead of the conservation of helicity and circulation. This raises the question of a ``classical limit'' in which a relevant notion of helicity is recovered which has dynamics akin to helicity in classical flows. To answer this question, we study bundles of superfluid vortices that mimic the structure of classical vortices and are robust long-lived structures \cite{alamri_reconnection_2008,wacks_coherent_2014}. Our numerical simulations show that the centerline helicity \cite{scheeler_helicity_2014} of superfluid vortex bundles behaves akin to helicity in classical viscous flows.

\section{Superfluid vortex dynamics and consequences for helicity}
\begin{figure}[!htb]
\includegraphics[width=0.7\columnwidth]{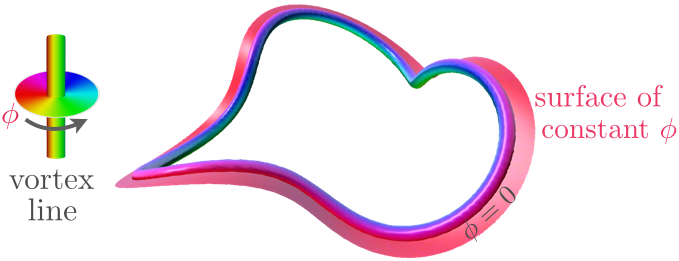}
\caption{A three-fold helical superfluid vortex and a section of its phase isosurface clipped at a fixed distance from the vortex. The volume occupied by the superfluid naturally separates into such surfaces of constant phase.}
\label{superfluid_vortex_evolve}
\end{figure}

To simplify our discussion, we consider superfluids at zero temperature, i.e.\ weakly interacting Bose condensates described by a complex order parameter $\psi$ (``wave function of the condensate'' \cite{dalfovo_theory_1999}) obeying the Gross-Pitaevskii equation \cite{gross_hydrodynamics_1963,pitaevskii_vortex_1961}:
\begin{equation}
 i \hbar\,\partial_t \psi = -\frac{\hbar^2}{2m} \nabla^2 \psi + g\,\vert \psi \vert^2 \,\psi \label{gpe}
\end{equation}
where the constant $g$ captures the inter-atomic interaction strength \cite{barenghi_vortices_2016}.
The Gross-Pitaevskii equation (GPE) captures qualitatively important features of superfluid behavior at low temperatures \cite{donnelly_quantized_1991,barenghi_quantized_2001}, including the dynamics of vortices---lines where the complex order parameter $\psi$ vanishes, and around which its phase winds around by an integer multiple of $2\pi$ (see Fig.~\ref{superfluid_vortex_evolve}).

Interestingly, the Gross-Pitaevskii equation can be mapped to an Euler flow in the region excluding vortices via the Madelung transformation \cite{madelung_anschauliche_1926,madelung_quantentheorie_1927}: $\psi=\sqrt{\rho/m}\exp(i\phi/\hbar)$, by rewriting Eq.~(\ref{gpe}) in terms of the fluid density $\rho=m\vert \psi \vert^2$, and velocity $\mathbf{u}=\nabla\phi/m\,$. The mapping between superfluid flow and Euler flow makes it tempting to conclude that classical helicity is conserved in superfluids just as in Euler flows. However, numerical simulations show that the expression for helicity in Euler flows: $\mathcal{H}_\textrm{Euler} = \int\textrm{d}^3x\, \mathbf{u}\cdot\,\pmb{\omega}\;,\; \pmb{\omega}=\nabla\times\mathbf{u}$ is not conserved in superfluid flows \cite{scheeler_helicity_2014,clark_di_leoni_helicity_2016,hanninen_helicity_2016}. $\mathcal{H}_\textrm{Euler}$ evaluated for singular vortex lines has two contributions: (a) the Gauss linking integral for pairs of vortex lines, giving the linking between them, and (b) the Gauss linking integral evaluated for each vortex line and itself giving its writhe \cite{deturck_linking_2013}. Since the writhe of a vortex line is not conserved \cite{scheeler_helicity_2014} even in the absence of reconnections, $\mathcal{H}_\textrm{Euler}$ is not conserved for superfluid flows.

This disparity between Euler flows and superfluid flows stems from two key differences: (i) Superfluids have singular vorticity distributions, concentrated on lines of singular phase (see Fig.~\ref{superfluid_vortex_evolve}), and quantized circulation $\Gamma = \oint \mathbf{u}\cdot \textrm{d}\mathbf{l} =n\,h/m$, unlike classical vortices which have smooth vorticity distributions. (ii) Vortex lines in a superfluid can reconnect \cite{koplik_vortex_1993,proment_vortex_2012,bewley_characterization_2008}, in contrast to vortex lines in Euler flows which can never cross.

The singular nature of superfluid vortices and the presence of vortex reconnections make it challenging to carry over the derivation of helicity conservation \cite{moffatt_degree_1969} in Euler flows, and suggest that a fundamentally different approach is required to address the question of a ``superfluid helicity''. Previous approaches \cite{hanninen_helicity_2016,bekenstein_conservation_1992,mendelson_cosmic_1999} to seeking a conserved quantity analogous to helicity in superfluid flows have focused on adapting the expression for classical helicity $\mathcal{H}_\textrm{Euler}$ to superfluids, as opposed to starting from a symmetry and seeking conservation laws.

We now begin with the fundamental symmetry that gives rise to helicity conservation in Euler flows via Noether's theorem, and carry this over to superfluids.

\section{Helicity as a Noether charge for Euler fluids and Superfluids}
The conservation of helicity in Euler flows \cite{marsden_introduction_1999,marsden_reduction_2000,morrison_hamiltonian_1998,padhye_fluid_1996,padhye_relabeling_1996,kuroda_symmetries_1990,fukumoto_unified_2008,salmon_hamiltonian_1988,cotter_noethers_2012,tao_noethers_2014} is a special conservation law, arising from the particle relabeling symmetry via Noether's second\footnote{For more details on Noether's second theorem, see \cite{olver_applications_1986,kosmann-schwarzbach_noether_2011}.} theorem \cite{padhye_relabeling_1996,webb_magnetohydrodynamics_2018}. The particle relabeling symmetry arises from an equivalence between the Lagrangian description of a flow in terms of the positions $\mathbf{x}(\mathbf{a},\tau)$ and velocities $\partial_\tau\mathbf{x}(\mathbf{a},\tau)$ of fluid particles labeled by $\mathbf{a}$ at time $\tau$, and the Eulerian description of a flow in terms of the velocity $\mathbf{u}(\mathbf{x},t)$ and density $\rho(\mathbf{x},t)$ at each point in space. The action for Euler flow is \cite{morrison_hamiltonian_1998,kuroda_symmetries_1990,salmon_hamiltonian_1988}:
\begin{equation}
S_\textrm{Euler} = \int d\tau\, d^3 a \left[ \frac{1}{2} \left(\partial_\tau\mathbf{x}(\mathbf{a},\tau)\right)^2 - E(\rho)  \right] 
\label{euler_action}
\end{equation}
where $\tau$ is time, $\textrm{d}^3a = \rho\,\textrm{d}^3x$ is the mass of a fluid element, $\partial_\tau \mathbf{x}(\mathbf{a},\tau)$ is the velocity, $E(\rho(\mathbf{a}))$ is the internal energy density, and the co-ordinate frames $(\mathbf{a},\tau)$ and $(\mathbf{x},t)$ are related as follows: $\partial_\tau = \partial_t + \mathbf{u}\cdot\nabla\,$. Note that the Euler flow action in Eq.~(\ref{euler_action}) depends only on the flow velocity $\mathbf{u} = \partial_\tau\mathbf{x}(\mathbf{a},\tau)$, and the density $\rho: \rho^{-1}(\mathbf{a})=\det\left({\partial x^i(\mathbf{a})/\partial \,a^j}\right)$.

Particle labels can be interpreted as the initial co-ordinates of the fluid particles, and the relabeling transformation as a smooth reshuffling (diffeomorphism) of the particle labels, akin to a passive co-ordinate transformation, which leaves the fluid velocity and density unaffected and hence leaves the action invariant.

Relabeling transformations are changes of the particle labels:
\( a^i \to \tilde{a}^i = a^i + \epsilon\, \eta^i,\) where $\eta^i$ satisfies: (i) \( \partial \eta^ i/\partial \tau = 0\) which ensures that the velocity is unchanged, and (ii) \(\partial \eta^i/\partial a^i=0\) which ensures that the density $\rho = \det\left(\partial\mathbf{x}/\partial\mathbf{a}\right)^{-1}$ is invariant. The positions of the fluid particles remain unchanged under such a transformation, i.e. $\tilde{\mathbf{x}}(\tilde{\mathbf{a}},\tau) = \mathbf{x}(\mathbf{a},\tau)$. The conserved charge associated with relabeling transformations \cite{kuroda_symmetries_1990,padhye_fluid_1996,padhye_relabeling_1996,morrison_hamiltonian_1998,fukumoto_unified_2008} is: 
\begin{equation}
\mathcal{Q}_\textrm{Euler} = \int d^3a\, u_i \,\frac{\partial x^i}{\partial a^j}\,\eta^j \label{noether_charge_euler}
\end{equation}
where $u_i = \partial x_i / \partial\tau$. 

The conservation of $\mathcal{Q}_\textrm{Euler}$ gives both Kelvin's circulation theorem, and helicity conservation for different choices of $\pmb\eta$. Evaluating $\mathcal{Q}_\textrm{Euler}$ for the relabeling transformation $\eta^j = \oint_{C:\mathbf{a}(s)}ds\, \delta^{(3)}(\mathbf{a}-\mathbf{a}(s))\,\partial a^j(s)/\partial s$ which infinitesimally translates particle labels along a loop $C$ \cite{kuroda_symmetries_1990,padhye_relabeling_1996,bretherton_note_1970} gives the circulation along the loop $C$: $\Gamma_{C}=\oint_{C}\mathbf{u}\cdot \textrm{d}\mathbf{x}(s)$. Evaluating $\mathcal{Q}_\textrm{Euler}$ for the relabeling transformation $\eta^j = \epsilon^{jkl} (\partial u_p/\partial a^k) (\partial x^p/\partial a^l)$ which infinitesimally translates the particle labels $\mathbf{a}$ along vortex lines, gives the helicity $\mathcal{H}_\textrm{Euler}=\int \mathbf{u}\cdot\omega\,\textrm{d}^3x$ \cite{kuroda_symmetries_1990,padhye_fluid_1996,padhye_relabeling_1996,morrison_hamiltonian_1998,fukumoto_unified_2008}. Conservation of helicity follows as a special case of Kelvin's circulation theorem: from the conservation of the sum of circulations \emph{along} all the vortex lines in the fluid, weighted by the flux of each vortex line.

\begin{figure}[!htb]
\includegraphics[width=\columnwidth]{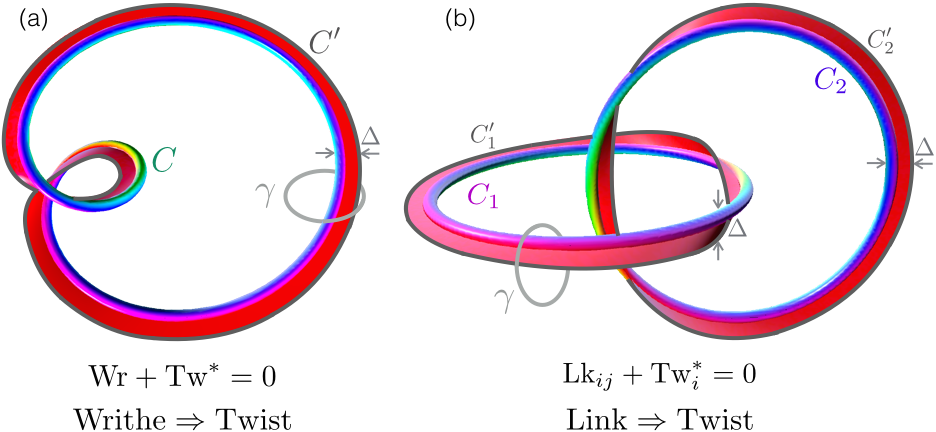}
\caption{Vortex lines $C$, and closed curves $C'$ constructed by offsetting vortex lines along a phase isosurface for: (a) a writhing (coiling) vortex line $C$, (b) a pair of linked rings $C_1\,,\,C_2$. Notice that the presence of either writhe or linking in vortex lines leads to the twisting of the phase isosurface around the vortex lines. The circulation around a closed loop $\gamma$ encircling a vortex line is equal to the change in phase $\phi$ as the loop is traversed, giving a multiple of $2\pi$.}
\label{twist_fig}
\end{figure}

We seek conserved quantities analogous to helicity and circulation in superfluids, by seeking analogs of the relabeling symmetry transformations. The action for the Gross-Pitaevskii superfluid in terms of the hydrodynamic variables $\rho = m\,\vert \psi\vert^2$, and $\phi = \hbar \arg\psi$ is: 
\begin{align*}
S_{\textrm{gpe}} &= - \int dt\,\rho\,d^3x\Big( \frac{\partial_t\phi}{m} + \frac{(\nabla\phi)^2}{2m^2} + \frac{g}{2m^2}\rho +\left(\frac{\hbar\, \nabla\sqrt{\rho}}{m\sqrt{2 \,\rho}}\right)^2 \Big)
\end{align*}
where the last term: $(\nabla\sqrt{\rho}/\sqrt{\rho})^2$ is known as the ``quantum pressure'' term, and has no classical analogue. Its primary effect is to regularize the size of the vortex core \cite{miniatura_ultracold_2011,rindler-daller_angular_2012,roberts_nonlinear_2001} and enable vortex reconnections \cite{barenghi_vortices_2016}, and is negligible when the typical length scale of density variations is much larger \cite{barenghi_vortices_2016}
than the ``healing length'' $\xi=\sqrt{\hbar^2/(2m\,g\,\rho_\textrm{max})}$. We make the Thomas-Fermi approximation \cite{barenghi_vortices_2016,pitaevskii_bose-einstein_2003,dalfovo_theory_1999} which neglects the ``quantum pressure'' term and captures well, the dynamics of superfluid vortices \cite{barenghi_vortices_2016,pitaevskii_bose-einstein_2003,jerrard_vortex_2002,lund_defect_1991}. Within this approximation, we seek to express the action for the Gross-Pitaevskii superfluid in terms of Lagrangian co-ordinates $(\mathbf{a},\tau)$, where $\mathbf{a}$ is the particle label, and $\tau$ is time. To this end, we rewrite $\nabla\phi$ as the fluid velocity $\nabla\phi/m=\mathbf{u}=\partial\mathbf{x}(\mathbf{a},\tau)/\partial\tau$, and use the relation $\partial_\tau = \partial_t + \mathbf{u}\cdot\nabla$ to rewrite $\partial_t\phi$ as $\partial_\tau\phi - \mathbf{u}\cdot\nabla\phi$. The superfluid action then becomes:
\begin{align*}
S_\textrm{gpe} &=  \int d\tau\, d^3 a \left[ \frac{1}{2}  \left(\partial_\tau\mathbf{x}(\mathbf{a},\tau)\right)^2 - E(\rho)  - \frac{1}{m}\partial_\tau\phi(\mathbf{a},\tau)\right] 
\end{align*}
where $E(\rho)=g\,\rho/(2m^2)$, $\rho\,\textrm{d}^3x = \textrm{d}^3a$ as for Euler flow. Note that the action $S_\textrm{gpe}$ differs from the Euler flow action in Eq.~(\ref{euler_action}) by an extra term: $\int d\tau\, d^3 a (- \partial_\tau\phi(\mathbf{a},\tau)/m)$. This extra term is needed to ensure Galilean invariance\footnote{as described in \cite{sulem_nonlinear_2004,kambe_elementary_2007}, under a Galilean transformation:$\{\mathbf{x}\rightarrow\mathbf{x}'=\mathbf{x}-\mathbf{v}t, t\rightarrow t'=t\}$, the phase transforms as: $\phi(\mathbf{x},t)\rightarrow \phi(\mathbf{x}',t)=\phi(\mathbf{x},t)-(\mathbf{v}\cdot\mathbf{x}-(\mathbf{v}\cdot\mathbf{v})t/2)$, assuming $m=\hbar=1$.}  of the action $S_\textrm{gpe}$, and has key consequences for the conservation of helicity.

Particle relabeling transformations of the form 
\( a^i \to \tilde{a}^i = a^i + \epsilon\, \eta^i\,,\,\tilde{\mathbf{x}}(\tilde{\mathbf{a}},\tau) = \mathbf{x}(\mathbf{a},\tau)\,,\,\tilde{\phi}(\tilde{\mathbf{a}},\tau) = \phi(\mathbf{a},\tau) \), where \( {\partial \eta^ i}/{\partial \tau} = 0\,,\, {\partial \eta^i}/{\partial a^i}=0 \), 
leave the velocity, the phase, and the density unchanged, and hence are symmetries of the action. Using Noether's theorem, the corresponding conserved charge is:
\begin{align}
\mathcal{Q}_\textrm{gpe} &= \quad\mathcal{Q}_\textrm{Euler} \qquad +\qquad  \mathcal{Q}_\textrm{phase} \nonumber \\
&= \int d^3a\, u_i \,\frac{\partial x^i}{\partial a^j}\,\eta^j +\int d^3a\, \left( \frac{-1}{m} \frac{\partial\phi}{\partial a^j}\right)\,\eta^j =0  \label{noether_charge_gpe}
\end{align}
where $\mathcal{Q}_\textrm{Euler}$ is the contribution from the Euler flow part of the action $S_\textrm{Euler}$, and $\mathcal{Q}_\textrm{phase} = \int d^3a \left(-\partial\phi/\partial a^j \right)\eta^j$ is the contribution from $S_\textrm{phase}$. The classical conserved charge $\mathcal{Q}_\textrm{Euler}$ is simply the superfluid conserved charge $\mathcal{Q}_\textrm{gpe}$ in the absence of $\mathcal{Q}_\textrm{phase}$ since the phase of the complex order parameter $\phi(\mathbf{a},\tau)$ is absent from the description of classical flow. Since the superfluid velocity is $\mathbf{u}=\nabla\phi/m$, $\mathcal{Q}_\textrm{Euler}$ and $\mathcal{Q}_\textrm{phase}$ cancel each other exactly. Hence, the conserved charge $\mathcal{Q}_\textrm{gpe}$ vanishes identically for all relabeling transformations, instead of giving conservation of helicity and circulation.

Our calculation shows that even in the absence of a ``quantum pressure'' term, the relabeling symmetry yields a vanishing conserved quantity, instead of conservation of circulation and helicity. This vanishing of ``superfluid helicity'' is consistent with an alternative calculation based on helicity as a Casimir invariant \cite{morrison_hamiltonian_1998,kuroda_symmetries_1990} (see SI for details). 

\section{Superfluid helicity---a geometric interpretation}
The vanishing of superfluid helicity and circulation $Q_{\rm gpe}$, is a consequence of a relation between the geometry of superfluid vortex lines and phase isosurfaces, as we now illustrate. 

For a relabeling transformation\footnote{$\pmb\eta_\gamma = \oint_{\gamma}ds\, \delta^{(3)}(\mathbf{a}-\mathbf{a}(s))\,\textrm{d} \mathbf{a}(s)/\textrm{d}s$, where $\mathbf{a}(s)\in \gamma$} along a closed loop $\gamma$ encircling a vortex line as shown in Fig.~\ref{twist_fig}, the vanishing of the conserved charge comes from a cancellation between the circulation $\oint_\gamma \mathbf{u}\cdot \textrm{d}\mathbf{l}$ and the change in phase $\oint_\gamma \left( -\nabla\phi\right)\cdot\textrm{d}\mathbf{l}$. We note, however, that by judiciously choosing the shape of the loop, so that it lies entirely on a phase isosurface as depicted in Fig.~\ref{twist_fig}, it is  possible to make the contribution $\mathcal{Q}_\textrm{phase}$ vanish identically. The vanishing of $\mathcal{Q}_\textrm{gpe}$ then acquires a simple geometric interpretation, which we elucidate below.

\begin{figure}[!htb]
\includegraphics[width=\columnwidth]{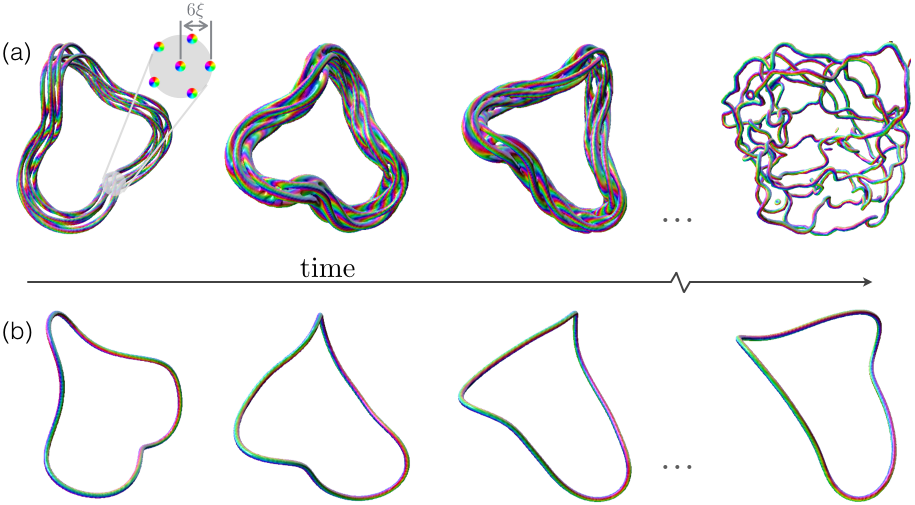}
\caption{A three-fold helical superfluid vortex bundle (shown in (a)) evolving as a coherent structure, rotating as it travels forward, akin to a single three-fold helical vortex (shown in (b)). A cross-section of the three-fold helical superfluid vortex bundle, reveals a central vortex and $5$ equally spaced vortices arranged around the central vortex at distance $6\xi$ (where $\xi$ is the healing length). After a long time, the helical vortex bundle disintegrates (symbolized by the grey dots) and loses its bundle-like structure.}
\label{superfluid_vortex_bundle_evolve}
\end{figure}

A curve along which $\mathcal{Q}_\textrm{phase}$ vanishes identically is constructed by offseting the vortex line $C_i$ along a phase isosurface by a distance $\Delta$ (see Fig.~\ref{twist_fig}) to give a new closed curve $C'_i(\Delta):\mathbf{a}'(s) = \mathbf{a}(s) +\Delta\,\hat{\mathbf{n}}(s)$, where $\mathbf{a}(s) \in C_i$, and $\hat{\mathbf{n}}(s)$ is perpendicular to the vortex line and tangent to the phase isosurface. The quantum pressure term is negligible on the new closed curve $C'_i(\Delta)$ as long as the distance $\Delta$ is large compared to the healing length $\xi$. The conserved charge $\mathcal{Q}_\textrm{gpe}$ evaluated for a relabeling transformation $\pmb\eta(\Delta)$\footnote{$\pmb\eta(\Delta) = \oint_{C'_i(\Delta)}ds\, \delta^{(3)}(\mathbf{a}-\mathbf{a}'(s))\,\textrm{d} \mathbf{a}'(s)/\textrm{d}s$} which translates particle labels along $C'_i(\Delta)$ has no contribution from $\mathcal{Q}_\textrm{phase}$, and becomes the circulation along the curve $C'_i(\Delta)$:  $\mathcal{Q}_\textrm{gpe}=\oint_{C'_i(\Delta)} \mathbf{u}\cdot \textrm{d}\mathbf{l}$. This circulation can be evaluated by substituting the Biot-Savart flow field for $\mathbf{u}$, since the compressible part of $\mathbf{u}$ does not contribute. 

$\mathcal{Q}_\textrm{gpe}$ then becomes the linking of the loop $C'_i$ with all the vortex lines in the superfluid, i.e. $\mathcal{Q}_{\rm gpe} =\sum_{j\neq i} \Gamma_j \,\mathcal{L}_{i'j} + \Gamma_i \,\mathcal{L}_{i'i} = 0$
where $\mathcal{L}_{i'j}$ denotes the linking between the vortex line $C_j$, and we have used the Gauss linking integral \cite{ricca_gauss_2011}. The vanishing of the conserved charge $\mathcal{Q}_\textrm{gpe}$ follows as result of the linking $\mathcal{L}_{i'i}$ between the offset line $C'_i$ and the vortex line $C_i$ canceling the linking $\mathcal{L}_{i'j}$ between the offset line $C'_i$ and all the other vortex lines $C_j\,,\, j\neq i$. Furthermore, assuming that the section of the phase isosurface bounded by the two loops $C'_i$, $C_i$ can be considered as a smooth ribbon, we can use the C\v{a}lug\v{a}reanu-White-Fuller theorem \cite{calugareanu_lintegrale_1959,calugareanu_sur_1961,white_self-linking_1969,fuller_writhing_1971} to express $\mathcal{L}_{i'i}$ as the sum of the writhe $(\textrm{Wr}_i)$ and the twist $(\textrm{Tw}^\ast_i)$ of the ribbon (see Fig.~\ref{twist_fig}), giving:
\begin{equation}
\mathcal{Q}_\textrm{gpe} = \sum_{j\neq i} \Gamma_j \mathcal{L}_{ij} + \Gamma_i \mathrm{Wr}_i + \Gamma_i \mathrm{Tw}^\ast_i = 0 \label{lk_tw_wr}
\end{equation}
The vanishing of the conserved charge $\mathcal{Q}_\textrm{gpe}$ is thus related to the vanishing of the sum of: the linking of a vortex line $C_i$ with all other vortex lines $\sum_{j\neq i}\mathcal{L}_{ij}$, its writhe $\mathrm{Wr}_i$, and the twist $\mathrm{Tw}^\ast_i$ of a ribbon formed by a phase isosurface ending on it. 

The vanishing of these geometric quantities was first studied in the context of helicity of framings of magnetic flux tubes \cite{akhmetev_borromeanism_1992}, and is a consequence of the fact that a phase isosurface is an orientable surface which has as its boundary, all the vortex lines in the superfluid, i.e. it is a Seifert surface \cite{seifert_uber_1935,akhmetev_borromeanism_1992,ruzmaikin_topological_1994,wijk_visualization_2006} for the vortex lines in the superfluid. This relation between linking and writhing of vortex lines and the twisting of phase isosurfaces has been used in superfluid simulations \cite{scheeler_helicity_2014,kleckner_how_2016} to calculate the centerline helicity (linking and writhing of vortex lines), and was elaborated on in recent efforts to define a superfluid helicity \cite{hanninen_helicity_2016,salman_helicity_2017}.

\begin{figure}[!htb]
\includegraphics[width=\columnwidth]{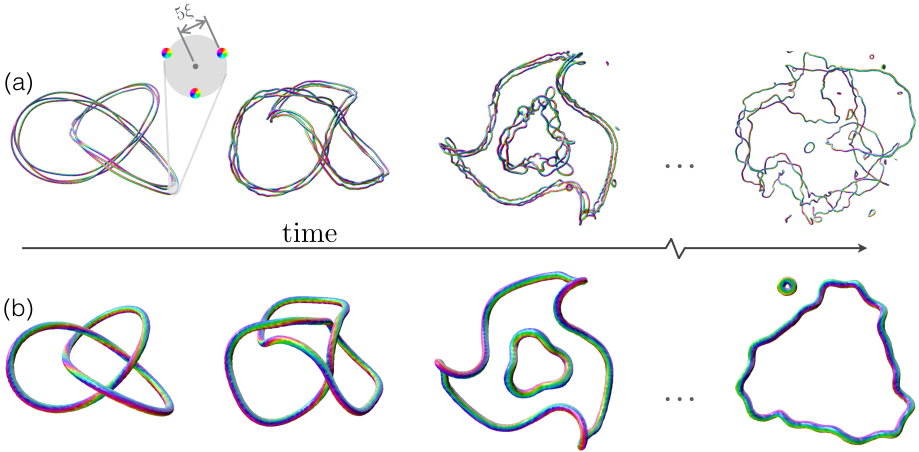}
\caption{A superfluid vortex bundle in the shape of a trefoil knot evolving as a coherent structure, akin to a single trefoil knot vortex. (a) A trefoil knotted vortex bundle reconnects to form a smaller three-fold distorted ring bundle, and a larger three-fold distorted ring bundle, which lose their bundle-like structure over time.  A cross-section of the initial trefoil knotted vortex bundle, shows $3$ equally spaced vortices arranged on the circumference of a disk of radius $5\xi$. (b) A single trefoil knotted vortex reconnects to form a smaller three-fold distorted ring, and a larger three-fold distorted ring, which undergoes further reconnections to give a large distorted ring at long times.}
\label{trefoil_bundle}
\end{figure}

\section{Classical helicity---the singular limit and dissipation}
We now address the question of whether a classical notion of helicity can be recovered in superfluids and if its dynamics are akin to that in Euler flows or viscous flows. 

\begin{figure*}[!hbt]
\includegraphics[width=2\columnwidth]{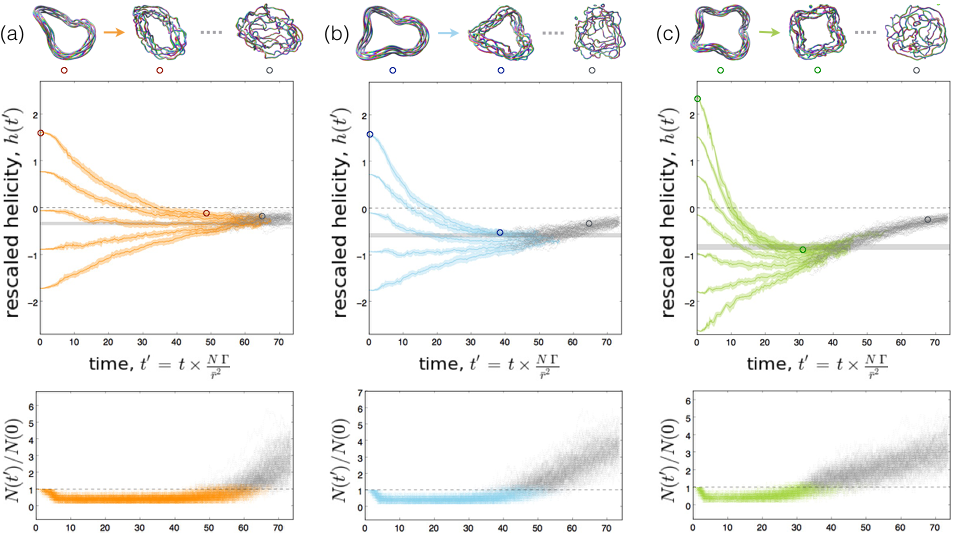}
\caption{Helical vortex bundles (N=6) at different stages of evolution (top row), with the corresponding points in the graphs indicated by colored circles (bundle-like structure preserved), and grey circles (bundles disintegrate). (a) 2-fold helical vortex bundles with aspect ratio 0.35, (b) 3-fold helical vortex bundles with aspect ratio 0.25, and (c) 4-fold helical vortex bundles with aspect ratio 0.2. The rescaled helicity $h$ (middle row) for superfluid vortex bundles having the same overall shape (writhe) but different amounts of twist, trends towards their initial average writhe (horizontal grey band), before eventually decaying towards zero (grey dotted lines). After a vortex bundle disintegrates at time T (=$\min t': N(t')/N(0) > 1.5$), its rescaled helicity is shown by a grey dotted line. Bottom panel shows the ratio of the number of vortex filaments at time $t'$ to the initial number of vortex filaments: $N(t')/N(0)$. For each helical vortex bundle configuration, multiple ($>10$) simulations are performed with random Gaussian noise (r.m.s is $2\%$ of the r.m.s. radius) added to the initial bundle. The mean rescaled helicity is indicated by the solid lines, and the width of the shaded band around the solid line indicates the standard deviation ($2\sigma$).}
\label{twist_diss}
\end{figure*}

While vorticity in superfluids is necessarily concentrated on lines of singular phase, vorticity in classical fluids can be continuously distributed and indeed must be to avoid a physical singularity in the flow. Following \cite{moffatt_degree_1969,arnold_topological_1998,berger_introduction_1999}, a natural way of recovering a ``classical'' notion of helicity  is to consider a continuous vorticity distribution as made up of an infinite collection of vortex lines. The centerline helicity $\mathcal{H}_c$ of a collection of singular vortex lines is: 
\begin{align}
\mathcal{H}_c  &= \sum_{i}\sum_{j}\Gamma_i\Gamma_j \mathcal{L}_{ij} = \sum_i\sum_{j\neq i}\Gamma_i\Gamma_j \mathcal{L}_{ij} + \sum_i \Gamma_i^2 \mathcal{L}_{ii} \nonumber \\
&= \sum_i\sum_{j\neq i}\Gamma_i\Gamma_j \mathcal{L}_{ij} + \sum_i \Gamma_i^2\textrm{Wr}_i \label{singular_Hc}
\end{align}
where $\Gamma_i$ is the circulation around the $i^{th}$ vortex line, $\textrm{Wr}_i$ is the writhe of the $i^{th}$ vortex line, and $\mathcal{L}_{ij}$ is the linking between the $i^{th}$ and $j^{th}$ vortex lines. Since the above expression includes the writhe of vortex lines which is not a topological invariant, the centerline helicity of a collection of singular vortex lines is not conserved \cite{scheeler_helicity_2014}. Assuming that the circulation of each vortex line is $\Gamma$, the centerline helicity rescaled by the square of the total circulation $(N\,\Gamma)^2$ becomes:
\begin{align}
\frac{\mathcal{H}_c}{(N\,\Gamma)^2}  &= \frac{1}{N^2}\sum_i\sum_{j\neq i}\mathcal{L}_{ij} + \frac{1}{N^2}\sum_i \textrm{Wr}_i \label{rescaled_Hc}
\end{align}
In the limit $N\to\infty$, the contribution from the writhe term in Eq.~(\ref{rescaled_Hc}) scales as $O(1/N)$ and becomes irrelevant, as was shown in \cite{berger_topological_1984}, leaving only the contribution from the linking $\mathcal{L}_{ij}$ between different vortex lines which is conserved in Euler flows:  
\begin{equation}
\lim_{N\to\infty}\frac{\mathcal{H}_c}{(N\,\Gamma)^2} = \lim_{N\to\infty} \frac{1}{N^2}\sum_i\sum_{j\neq i} \mathcal{L}_{ij} = \frac{\mathcal{H}_\textrm{Euler}}{\Gamma^2_\textrm{total}} \label{Hc_infty}
\end{equation}
Hence the rescaled centerline helicity of an infinite collection of vortex lines is conserved in Euler flows. However, for a finite number of singular vortex lines, the writhe term remains relevant albeit $O(1/N)$ and the rescaled centerline helicity is not conserved. The case of a superfluid is interesting in the context of this discussion, since quantization imposes a fundamental granularity in the vorticity field. 

Since the above calculation is independent of the dynamics of the vortices, 
it leaves unanswered the question of what the dynamics of the rescaled centerline helicity of collections of superfluid vortex lines will be. In particular, will the centerline helicity remain unchanged as in Euler flows, follow the dynamics observed in viscous flows, or have entirely different dynamics?

In the case of Euler flows, the helicity dynamics are simple: $\mathcal{H}_c$ remains constant (in the limit of an infinite number of vortex lines). In the case of viscous flows, the dynamics are more subtle. For a freely evolving helical vortex, as shown in a recent study \cite{scheeler_complete_2017}, the total helicity converges to the writhe over time. This can be rationalized by separating the helicity into contributions from (a) the linking between bundles, (b) the writhing (coiling) of bundles and (c) the local twisting of vortex lines, with the total twist being the difference between the total helicity and the former two. Since the twist is the only local component of helicity, it is the only one acted upon by viscosity and thus the only one that dissipates.

The special role of twist can be understood by computing the instantaneous rate of helicity dissipation: $\partial_t\mathcal{H}=-2\nu\int \textrm{d}^3x\;\pmb{\omega}\cdot\nabla\times\pmb{\omega}=-2\nu\int \textrm{d}^3x\;\vert\pmb\omega\vert^2\,\hat{\pmb{\omega}}\cdot\nabla\times\hat{\pmb{\omega}}$, where $\hat{\pmb\omega}\cdot\nabla\times\hat{\pmb\omega}$ captures the local twisting of vortex lines \cite{kedia_construction_2017}, and vanishes identically for a twist-free thin-core vortex \cite{scheeler_complete_2017}. While the role of the twist-free state as the zero-dissipation state is clear, the dynamics of the approach to such a state are more challenging to study because of their dependence on the local details of the vortex core \cite{scheeler_complete_2017}.

Thus for a collection of superfluid vortices, a constant rescaled centerline helicity would suggest Euler-flow like behavior, while the convergence of the rescaled centerline helicity to the writhe would suggest viscous flow-like behavior. 

\begin{figure}[!hbt]
\includegraphics[width=\columnwidth]{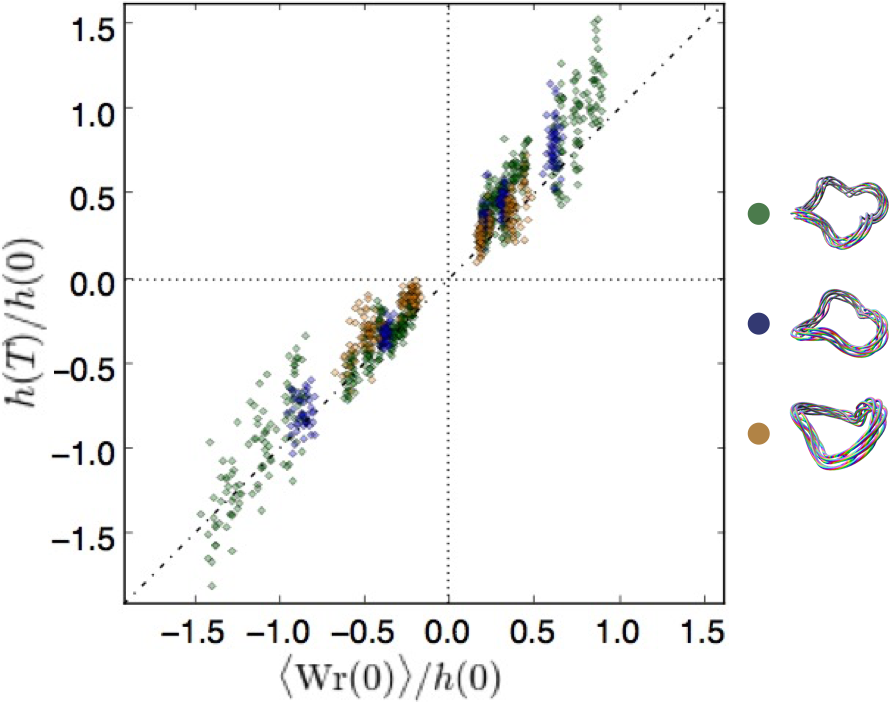}
\caption{The ratio $h(T)/h(0)$ approaches the ratio $\langle\textrm{Wr}(0) \rangle/h(0)$ of the average initial writhe to the initial rescaled helicity for a variety of helical vortex bundles (1209 simulations) in the shape of 2 (aspect ratio:0.35),3 (aspect ratio: 0.25), and 4-fold (aspect ratios: 0.16, 0.18, 0.2) helices with $N=5$ and $N=6$ vortex filaments where $T$ is a proxy for the time at which the vortex bundle disintegrates. To divide by the initial helicity $h(0)$, we only consider vortex bundles whose initial helicity satisfies: $\vert h(0) \vert > 0.25$. Vortex bundles with initial helicity $\vert h(0) \vert < 0.25$ also display similar behavior with $h(T)\to \langle\textrm{Wr}(0)\rangle$ as shown in Fig.~\ref{twist_diss} and the SI.}
\label{H_vs_Wr0}
\end{figure}

\section{Centerline helicity of superfluid vortex bundles}
Superfluid vortex bundles which approximate the structure of a classical thin-core vortex tube, have been shown to be robust coherent structures \cite{alamri_reconnection_2008,wacks_coherent_2014}. We construct thin bundles of equally spaced vortex lines winding around a central vortex loop as shown in Fig.~\ref{superfluid_vortex_bundle_evolve}(a), whose shape controls the writhe (coiling) of the vortex bundle. These superfluid vortex bundles evolve coherently over distances of the order of their size (see Figs.~\ref{superfluid_vortex_bundle_evolve},\ref{trefoil_bundle}, supplementary movies) before becoming unstable and disintegrating, as observed in previous work \cite{alamri_reconnection_2008,wacks_coherent_2014}. The coherent portion of the evolution of these bundles resembles the dynamics of single vortex loops in superfluids and the evolution of vortices in classical fluids, and has been studied for ring bundles \cite{wacks_coherent_2014} and reconnecting line bundles \cite{alamri_reconnection_2008}. When the vortex bundles become unstable, the number of individual vortices quickly proliferates as shown in the bottom panel of Fig.~\ref{twist_diss}, with the number of vortex strands acting as a natural indicator of whether the bundle has disintegrated. We use the earliest time $T$ at which the number of vortex filaments $N(T)$ exceeds their initial number $N_0$  by $50\%$ as the time until which the bundle evolves coherently. Figure \ref{twist_diss} shows that the transition between the coherent phase and the disintegration phase of the vortex bundle is sharp.

In order to inject different amounts of centerline helicity in the bundle, we twist\footnote{The twisting of vortex lines mentioned here describes the winding of one vortex line around another, and is distinct from the twist $\mathrm{Tw}^\ast$ in Eq.~(\ref{lk_tw_wr}) of the ribbon formed by a phase isosurface ending on a vortex line.} the lines of the bundle around the central vortex, thus varying the centerline helicity independently of the writhe of the bundle. An initial complex order parameter $\psi$ for these vortex bundles is constructed following the methods outlined in \cite{proment_vortex_2012,scheeler_helicity_2014,kleckner_how_2016}, and evolved by numerically solving the Gross-Pitaevskii equation (Eq.~(\ref{gpe})) using a split-step method. Simulations of vortex bundles in the shape of helices and trefoil knots show that their coherent evolution is much like their classical vortex tube counterparts \cite{kleckner_creation_2013,scheeler_helicity_2014}. Helical vortex bundles propagate coherently without a significant change in shape (see Fig.~\ref{superfluid_vortex_bundle_evolve}) for longer times, while knotted vortex bundles stretch and reconnect (see Fig.~\ref{trefoil_bundle}) to form disconnected loop bundles which quickly become unstable. Vortex bundles which evolve coherently over long times allow us to study the dynamics of their rescaled centerline helicity $h=\mathcal{H}_c/(N\,\Gamma)^2$. We focus on helical vortex bundles which evolve coherently over distances of $6\bar{r}$ or greater, and in particular study bundles in which the central vortex is a toroidal helix (see Figs.~\ref{twist_diss},\ref{H_vs_Wr0}) winding $2$,$3$,$4$ times in the poloidal direction around tori of aspect ratios $0.35$ (2-fold), $0.25$ (3-fold), $0.16,0.18,0.2$ (4-fold), as it winds around once in the toroidal direction. We consider superfluid vortex bundles with $N=5$ and $N=6$ vortex lines each having a circulation $\Gamma=2\pi$, an initial inter-vortex spacing of $d\sim 6\xi$ (see Fig.~\ref{superfluid_vortex_bundle_evolve}) and an overall r.m.s. radius $\bar{r}\sim 50\xi$. To avoid the possibility that symmetry stabilizes the vortices, we add a small amount of Gaussian noise to each vortex line in the transverse direction. To obtain sufficient statistics, we simulated the evolution of a total of 1,156 vortex bundles with a volume of $(256\xi)^3$ and a grid spacing of $1\xi$. A small number of simulations at double resolution (but the same physical volume) yield identical observations.

Unlike in Euler flows, where the rescaled centerline helicity $h$ of a bundle of singular vortex lines emerges as a conserved quantity in the limit of large $N$, the rescaled centerline helicity $h$ of superfluid vortex bundles appears to change with time. Assuming these superfluid vortex bundles approximate thin-cored vortex tubes, we can further decompose their rescaled centerline helicity (Eq.~(\ref{rescaled_Hc})) into contributions from the twisting of the vortex lines around each other, and their individual writhes. Using $\mathcal{L}_{ij}=\textrm{Tw}_{ij} +\textrm{Wr}_{i(j)}$, the rescaled centerline helicity becomes:
\begin{align}
\frac{\mathcal{H}_c(t)}{(N\,\Gamma)^2}  &= \frac{1}{N^2}\sum_i\sum_{j\neq i}\left(\textrm{Tw}_{ij}(t) + \textrm{Wr}_i(t) \right) + \frac{1}{N^2}\sum_i \textrm{Wr}_i(t) \nonumber \\
&= \frac{1}{N^2}\sum_i\sum_{j\neq i}\textrm{Tw}_{ij}(t) + \frac{1}{N}\sum_i \textrm{Wr}_i(t) \nonumber \\
&= \frac{1}{N^2}\sum_i\sum_{j\neq i}\textrm{Tw}_{ij}(t) + \langle \textrm{Wr}(t) \rangle \label{rescaled_Hc_Tw_Wr}
\end{align}
where the average writhe $\langle \textrm{Wr}(t) \rangle = \sum_i \textrm{Wr}_i(t)/N$ includes contributions from the writhe term in Eq.(\ref{rescaled_Hc}), as well as from the linking term by decomposing it into writhe and twist contributions. 

Our numerical simulations show that the rescaled centerline helicity of long-lived superfluid vortex bundles tends towards their average initial writhe $\langle \textrm{Wr}(0)\rangle$, as in Fig.s~\ref{twist_diss}, \ref{H_vs_Wr0}, suggesting\footnote{the difficulty of calculating the average writhe at later times stems from the small-wavelength fluctuations in the vortex lines which contribute to large fluctuations in their writhe.} that the twist term in Eq.~(\ref{rescaled_Hc_Tw_Wr}) decays over time. The dynamics of the rescaled centerline helicity $h$ are thus classical. 

The role of writhe in the dynamics of centerline helicity of superfluid vortex bundles in our simulations has a striking resemblance to the role of writhe in the helicity dynamics of vortices in viscous flows \cite{scheeler_complete_2017}. This points to a ``classical limit'' in which classical behavior is recovered from quantized vortex filaments \emph{geometrically} by replacing single vortex filaments with vortex bundles. However, owing to reconnections, the classical behavior that is recovered is not that of Euler flows, but that of the Navier-Stokes equations in which viscosity acts to dissipate twist. Our results corroborate the role of writhe as an attractor for the helicity at long times, adding a geometric lens to previous work \cite{barenghi_is_2008,clark_di_leoni_dual_2017} on the dissipative effects of vortex reconnections in superfluids.

\section{Conclusion}
We have addressed the existence of an additional conservation law in superfluids---conservation of helicity---by generalizing to superfluids the particle relabeling symmetry, which underlies helicity conservation in Euler flows. The application of Noether's second theorem to the particle relabeling symmetry \cite{padhye_relabeling_1996,webb_magnetohydrodynamics_2018} yields the conservation of helicity and circulation in Euler flows, however for superfluid flows it yields a trivially vanishing conserved quantity. 
This is owing to the appearance of an additional term that comes from the phase of the superfluid order parameter, not present in Euler flows. This additional term has a well-known geometric interpretation for the vanishing of ``superfluid helicity'' in terms of a relation between the linking and writhing of vortex lines, and the twisting of phase isosurfaces near vortex lines.

On replacing superfluid vortices with superfluid vortex \emph{bundles}, their centerline helicity becomes the classical helicity in the limit of an infinite collection of vortices. We study the dynamics of the centerline helicity of superfluid vortex bundles via numerical simulations and find behavior akin to that of classical helicity in a viscous fluid, with the writhe acting as an attractor for the final value of helicity.

\bibliography{refs}

\appendix

\section{Helicity---as a Casimir invariant}
In Euler flows, helicity emerges a special constant of motion: a Casimir invariant \cite{morrison_hamiltonian_1998,kuroda_symmetries_1990}, i.e. it has a vanishing Poisson bracket with any function of the phase space variables: \( \left\{ \mathcal{H}, F(\mathbf{u},\rho) \right\} = 0\,\;\forall \; F \), where the density $\rho$ and the fluid velocity $\mathbf{u}$ are the phase space variables, and $\left\{\, \cdot\,,\cdot\, \right\}$ denotes the Poisson bracket.
 
 Solving for the Casimir invariants in Euler flow, i.e. solving \( \left\{ \mathcal{C}, F(\mathbf{u},\rho) \right\} = 0\,\;\forall \; F \) gives rise to helicity as an additional conserved quantity. We seek an analogous conserved quantity in superfluids by solving for the Casimir invariants for the Gross-Pitaevskii equation. 
 
 The Hamiltonian corresponding to the Gross-Pitaevskii equation is:
 \begin{equation}
 \mathbb{H} = \int d^3 x\, \left[ \frac{\hbar^2}{2m}\vert \nabla \psi \vert^2 + \frac{V}{2}\vert \psi \vert^4  \right] \label{superfluid_hamiltonian}
 \end{equation}
 with the canonical Poisson bracket: 
 \begin{equation}
  \left\{ F\,,\, G \right\} =-\frac{i}{\hbar}\int d^3 x \left( \frac{\delta F}{\delta \psi } \frac{\delta G}{\delta \psi^\ast} - \frac{\delta G}{\delta \psi } \frac{\delta F}{\delta \psi^\ast} \right) 
 \label{superfluid_pb}
\end{equation}
Solving for the Casimir invariants $\left\{ \mathcal{C}, F(\psi,\psi^\ast) \right\} = 0\,\;\forall\; F$ reduces to the equations:
\begin{equation}
\frac{\delta\mathcal{C}}{\delta\psi} = 0 \; , \quad\frac{\delta\mathcal{C}}{\delta\psi^\ast} = 0 \label{gpe_casimir}
\end{equation}
 which gives only trivial constants as Casimir invariants. Since Casimir invariants of the Gross-Pitaevskii superfluid should yield a conserved quantity analogous to helicity in Euler flows, the above calculation suggests that the conserved  quantity analogous to helicity in superfluids is a trivial constant. This is consistent with our calculation based on the relabeling symmetry which suggests that the conserved quantity analogous to helicity in superfluids vanishes identically.

We note that an alternative path to seeking Casimir invariants, by taking the phase space variables to be $\left\{ \mathbf{j}\,, \rho \right\} = \left\{ \left( \psi^\ast \nabla\psi - \psi \nabla\psi^\ast\right)/(2i) \,,\, \psi^\ast\psi  \right\}$ instead of $\left\{\psi,\psi^\ast \right\}$ runs into difficulties because of the singular nature of vorticity: $\nabla\times \left(\mathbf{j}/\rho \right)$. This difficulty manifests in terms of an erroneous equation of motion for a vortex, as the Poisson bracket for the new phase space variables denoted by $\left\{\cdot\,,\,\cdot \right\}_{\mathbf{j},\rho}$ incorrectly gives: $\partial_t\left(\nabla\times \left( \mathbf{j}/\rho\right)\right) = \left\{ \nabla\times\left(\mathbf{j}/\rho\right)\,,\,\mathbb{H} \right\}_{\mathbf{j},\rho} = 0$, suggesting that vortex lines are stationary.

We now briefly review the underlying symmetry---the relabeling symmetry---that gives rise to helicity as a conserved charge via Noether's theorem, and calculate the analogous conserved charge in superfluids.

\begin{figure}[!hbt]
\includegraphics[width=\columnwidth]{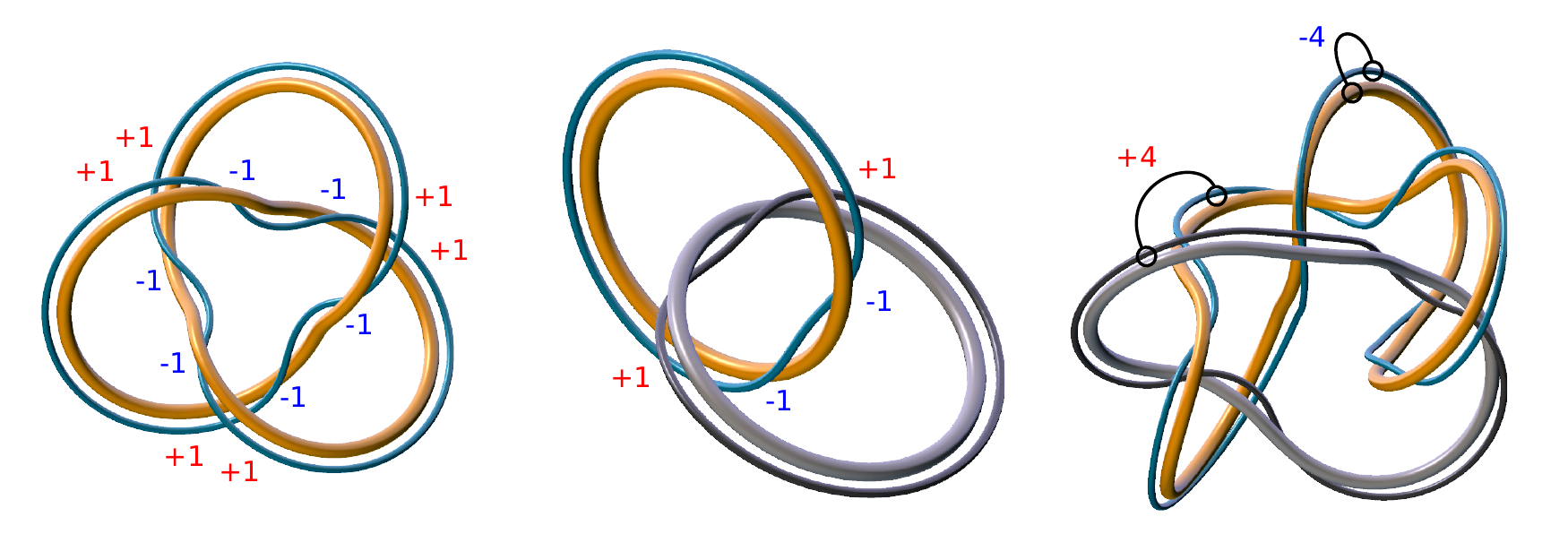}
\caption{An illustration of the linking of a vortex tangle with a path that is phase-offset from the original vortex. The orange
(light gray) lines indicate a vortex path, while the teal (dark gray) lines indicate a path offset from each vortex path along
a direction of constant phase. (The phase fields are computed by numerical integration of the Biot-Savart law.) Each signed
crossing of an offset line with the original vortex path is indicated; in each case the total linking between the offset path and
all vortex paths is 0. From left to right, the vortex topologies correspond to a trefoil knot, Hopf link, and $6^2_3$ from the Rolfsen table of links.}
\label{zero_link}
\end{figure}

\begin{figure*}[!hbt]
\includegraphics[width=2\columnwidth]{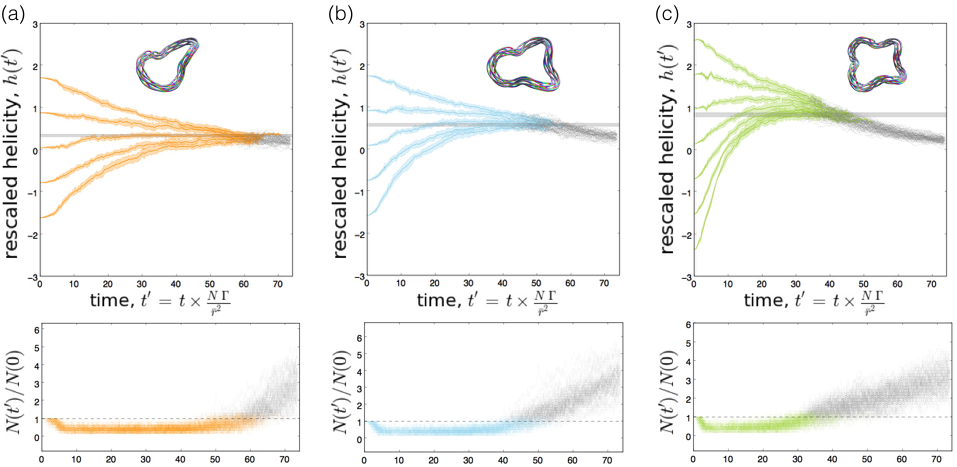}
\caption{Left-handed helical vortex bundles with positive initial writhe display helicity dynamics similiar to the right-handed helical vortex bundles shown in Fig.~5 in the main text. The rescaled helicity $h$ for superfluid vortex bundles constructed with varying degrees of twist, i.e. having different initial helicity, trends towards their initial average writhe (horizontal grey band) as long as the bundle-like structure is preserved, before eventually decaying towards zero (as indicated by the grey dotted lines) for (a) 2-fold helical vortex bundles, (b) 3-fold helical vortex bundles, and (c) 4-fold helical vortex bundles. For each helical vortex bundle configuration corresponding to a given initial rescaled helicity $h(0)$, multiple simulations are performed with random Gaussian noise (r.m.s is $2\%$ of the r.m.s. radius) added to the initial bundle. The mean rescaled helicity is indicated by the solid lines, and the width of the shaded band around the solid line indicates the standard deviation ($2\sigma$). After the vortex bundle disintegrates, its rescaled helicity is shown by a grey dotted line. The bottom row shows the ratio of the number of vortex filaments at time $t'$ to the initial number of vortex filaments: $N(t')/N(0)$. The time at which a vortex bundle disintegrates is measured as the earliest time at which the number of vortex filaments $N(t')$ exceeds the initial number of vortex filaments $N(0)$ by more than $50\%$.}
\label{twist_diss_si}
\end{figure*}

\begin{figure}[!hbt]
\includegraphics[width=\columnwidth]{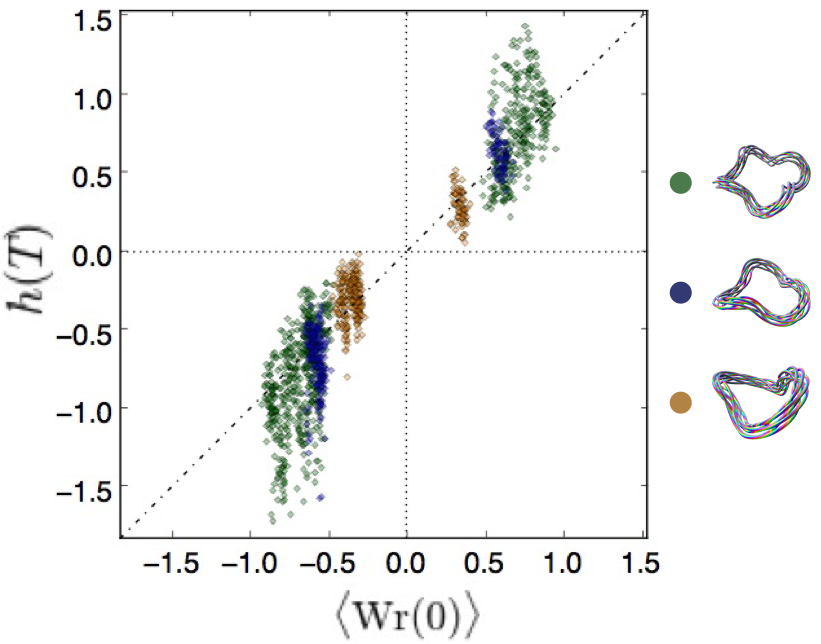}
\caption{The rescaled helicity $h(T)$ trends towards the average initial writhe $\langle\textrm{Wr}(0) \rangle$ for a variety of helical vortex bundles in the shape of 2,3, and 4-fold helices with $N=5$ and $N=6$ vortex filaments. Here $T$ is the time at which the vortex bundle disintegrates, i.e. the earliest time at which the number filaments $N(t')$ exceeds the initial number of filaments $N(0)$ by more than $50\%$. The large spread in values of $h(T)$ comes from vortex bundles whose initial rescaled helicity $h(0)$ is far from their average initial writhe $\langle \textrm{Wr}(0) \rangle$, and is removed on rescaling both the axes by $h(0)$, as shown in Fig.~6 in the main text. The final rescaled helicity $h(T)$ trends towards the average initial writhe as shown in Figs.~\ref{twist_diss},\ref{twist_diss_si} but such vortex bundles often disintegrate before the final rescaled helicity $h(T)$ becomes equal to the average initial writhe $\langle \textrm{Wr}(0) \rangle$, giving rise to the large observed spread in $h(T)$.}
\label{H_vs_Wr0_si}
\end{figure}

\begin{figure}[!hbt]
\includegraphics[width=\columnwidth]{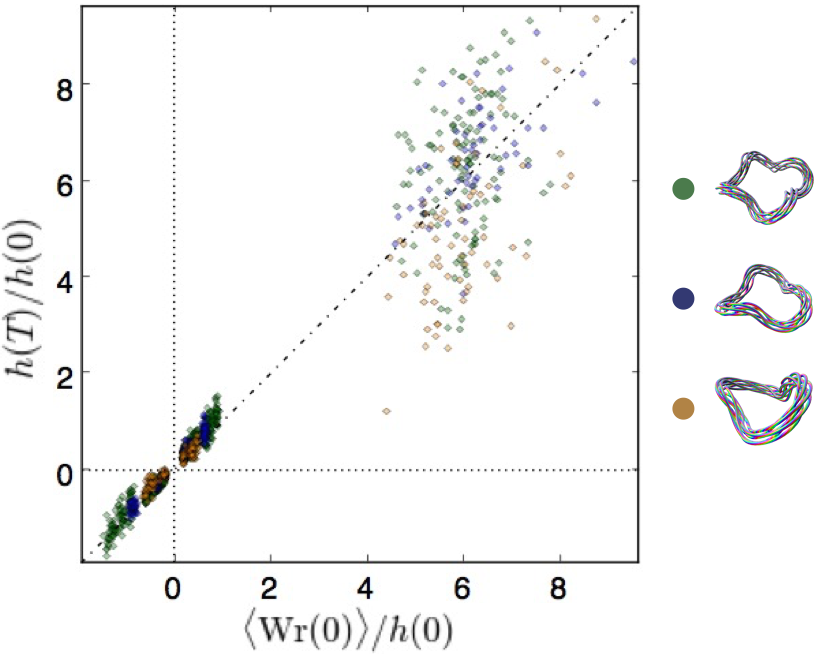}
\caption{Fig.~6 of main text including vortex bundles with lower initial helicity, i.e. $\vert h(0) \vert < 0.25$. The larger spread comes from dividing by a small number i.e. $h(0)$. The ratio of the rescaled helicity $h(T)$ to the initial rescaled helicity $h(0)$ approaches the ratio of the average initial writhe $\langle\textrm{Wr}(0) \rangle$ to the initial rescaled helicity for a variety of helical vortex bundles in the shape of 2,3, and 4-fold helices with $N=5$ and $N=6$ vortex filaments. Here $T$ is the time at which the vortex bundle disintegrates, i.e. the earliest time at which the number filaments $N(t')$ exceeds the initial number of filaments $N(0)$ by more than $50\%$.}
\label{h_by_h0_vs_Wr0_by_h0}
\end{figure}

\section{Helicity as a Noether charge}
We now consider a classical fluid which obeys the same equation of motion as the Gross-Pitaevskii superfluid, except for the quantum pressure term, and show that the relabeling symmetry which gives rise to helicity conservation via Noether's theorem in Euler fluids gives a vanishing conserved charge in such a classical fluid.

\subsection{Superfluid equations of motion}
On setting $\hbar = m = 1$, the Gross-Pitaevskii equation of motion for a superfluid is: 
\begin{equation}
i\partial_t \psi = -\frac{1}{2}\nabla^2\psi + V \vert\psi\vert^2\, \psi \label{gpe_superfluid}
\end{equation}
On substituting $\psi = \sqrt{\rho}\exp(i\phi)$, the above complex equation gives two real equations for the evolution of $\rho$ and $\phi$ as follows:
\begin{align}
\partial_t\rho &+ \nabla\cdot\left(\rho\,\nabla\phi \right) = 0 \label{gpe_mass_cons} \\
\partial_t\phi &+ \frac{1}{2}\left(\nabla\phi \right)^2 + V\rho - \frac{1}{2}\left(\frac{\nabla^2\sqrt{\rho}}{\sqrt{\rho}} \right) = 0 \label{gpe_phase}
\end{align}
On applying a spatial gradient operator $\nabla$ to Eq.~(\ref{gpe_phase}), and substituting the expression for the superfluid velocity $\mathbf{u}=\nabla\phi$, we find: 
\begin{equation}
\partial_t\mathbf{u} + \nabla\left( \frac{1}{2}\mathbf{u}^2 + V\rho - \frac{\nabla^2\sqrt{\rho}}{2\sqrt{\rho}}\right) = 0\label{gpe_euler} 
\end{equation}
Note that the above equation contains the quantum pressure term $\tfrac{1}{2}\nabla\left(\nabla^2\sqrt{\rho}/\sqrt{\rho} \right)$ containing spatial derivatives of the density, is dominant only near the vortex core. Such a term is not present in classical hydrodynamics, since the pressure is assumed to depend only on the local density, and not on the spatial derivatives of the density.  
We now make the Thomas-Fermi approximation \cite{dalfovo_theory_1999,kaiser_coherent_2001,wang_cold_2001,kaiser_coherent_2001,pitaevskii_bose-einstein_2003} and neglect the quantum pressure term  in the above equation, thereby considering a hypothetical classical fluid which obeys the above equation of motion without the quantum pressure term, i.e.
\begin{equation}
\partial_t\mathbf{u} + \nabla\left( \frac{1}{2}\mathbf{u}^2 + V\rho \right) = 0 \label{gpe_madelung}
\end{equation}
The above equation describes the superfluid well in the region excluding the vortex core. Note that the above equation is similar to the equation of motion for an irrotational Euler fluid:
\begin{equation}
\partial_t\mathbf{u} + \nabla\left(\frac{1}{2}\mathbf{u}^2 + e \right) = 0 \label{euler_irrotational}
\end{equation}
where $e\,:\, de = dp/\rho$ is the enthalpy per unit mass and p is the pressure.

\subsection{Relabeling symmetry in a classical Euler fluid}
The action for a classical (isentropic) Euler fluid is:
\begin{equation}
S_{\textrm{Euler}} = \int d^3 a\, d\tau \left( \frac{1}{2}\left(\frac{\partial\mathbf{x}(\mathbf{a},\tau)}{\partial\tau}\right)^2 - E(\rho) \right) 
\end{equation}
where $\mathbf{x}(\mathbf{a},\tau)$ is the position of the fluid element labeled by $\mathbf{a}$ at time $\tau$, and the fluid velocity $\mathbf{u}(\mathbf{a},\tau) = \partial_\tau\mathbf{x}(\mathbf{a},\tau)$. The label co-ordinates $\mathbf{a}$ are chosen such that $\rho\,d^3x = d^3 a\Rightarrow \tfrac{\partial(\mathbf{x})}{\partial(\mathbf{a})} = \rho^{-1}$. It is easily verified \cite{padhye_fluid_1996,kuroda_symmetries_1990,morrison_hamiltonian_1998} that extremizing the action with respect to variations in the position field $\mathbf{x}(\mathbf{a},\tau)$, gives the Euler equations of motion. Mass conservation follows from: $\tfrac{\partial}{\partial\tau}\rho^{-1} = \tfrac{\partial}{\partial\tau}\left( \tfrac{\partial(\mathbf{x})}{\partial(\mathbf{a})}\right)$. 

As shown in \cite{kuroda_symmetries_1990,padhye_fluid_1996,morrison_hamiltonian_1998,fukumoto_unified_2008} and can be easily verified, the transformation $a^i\rightarrow \tilde{a}^i = a^i + \epsilon\,\eta^i$, such that $\tfrac{\partial}{\partial\tau}\eta^i = 0, \tfrac{\partial}{\partial a^i}\eta^i=0$ is a symmetry of the action and gives the corresponding conserved Noether charge: 
\begin{equation}
\mathcal{Q} = \int d^3a\, u_i \,\tfrac{\partial x^i}{\partial a^j}\,\eta^j \label{noether_Q_euler}
\end{equation}
When the fluid labels are displaced infinitesimally along a closed material curve, the conserved charge $\mathcal{Q}$ simplifies to the circulation around the material loop $\Gamma_{C}$, thus giving Kelvin's circulation theorem. This can be verified by substituting $\eta^j = \oint_{C:\mathbf{a}(s)}ds\, \delta^{(3)}(\mathbf{a}-\mathbf{a}(s))\,\tfrac{\partial a^j(s)}{\partial s}$ in Eq.~(\ref{noether_Q_euler}). 
When the fluid labels are displaced infinitesimally along vortex lines, the conserved charge $\mathcal{Q}$ is the helicity of the fluid: $\mathcal{Q} = \mathcal{H} = \int d^3x \, \mathbf{u}\cdot\nabla\times\mathbf{u}$. This can be verified by substituting $\eta^j = \epsilon^{jkl} \tfrac{\partial}{\partial a^k} u_p\, \tfrac{\partial}{\partial a^l} x^p$ in Eq.~(\ref{noether_Q_euler}).

\subsection{Relabeling symmetry in a superfluid}
The action corresponding to the Gross-Pitaevskii equation is: 
\begin{equation}
S_{\textrm{gpe}} = \int dt\,d^3x\left( i\psi^\ast\partial_t\psi - \frac{1}{2}\vert\nabla\psi\vert^2 -\frac{V}{2}\vert\psi\vert^4 \right) \label{gpe_action_wavefn}
\end{equation}
which can be written in terms of $\rho,\phi$ as follows:
\begin{align}
S_{\textrm{gpe}} &= - \int dt\,d^3x\Bigg( \rho\, \partial_t\phi + \frac{1}{2}\rho\,(\nabla\phi)^2 + \frac{V}{2}\rho^2 +\frac{1}{2}(\nabla\sqrt{\rho})^2 \Bigg) \label{gpe_action_rho_phi}
\end{align}

It is easy to verify that extremizing the above action in Eq.~(\ref{gpe_action_rho_phi}) with respect to $\rho,\phi$ gives the desired equations of motion: Eq.s~(\ref{gpe_mass_cons}),(\ref{gpe_phase}), and that the last term in the action: $\tfrac{1}{2}(\nabla\sqrt{\rho})^2$ corresponds to the quantum pressure term in Eq.s~(\ref{gpe_phase}),(\ref{gpe_euler}). 

We now model the superfluid in the region excluding vortex cores as a classical fluid which carries with it a phase $\phi(\mathbf{x},t)$. We neglect the quantum pressure term (making the Thomas-Fermi approximation), and use the relation $\mathbf{u}=\nabla\phi$ to get the following new action:
\begin{align}
\tilde{S}_{\textrm{gpe}} &= - \int dt\,d^3x\left(  \frac{1}{2}\rho\,\mathbf{u}^2 + \rho\, \partial_t\phi +\frac{V}{2}\rho \right) \label{gpe_classical} 
\end{align}

In region excluding the vortex cores, we assume that we can label the fluid particles with labels $\mathbf{a}$ where $d^3a = \rho \,d^3 x$, and track the positions of these particles $\mathbf{x}(\mathbf{a},\tau)$ over time $\tau$. We now rewrite the above action in terms of label co-ordinates $\mathbf{a},\tau$ using $\partial_\tau = \partial_t + \mathbf{u}\cdot\nabla$: 

\begin{align}
\tilde{S}_{\textrm{gpe}} &=  \int d\tau\,d^3a\left( \frac{1}{2}\mathbf{u}^2 - \partial_\tau\phi -\frac{V}{2}\rho \right) \label{gpe_labels}
\end{align}

It is easy to verify that extremizing the above action with respect to $\mathbf{x}(\mathbf{a},\tau)$ gives the desired hydrodynamic equation of motion: Eq.~(\ref{gpe_madelung}), suggesting that transforming $S_\textrm{gpe}$ in Eq.~(\ref{gpe_action_rho_phi}) to the above action $\tilde{S}_{\textrm{gpe}}$ in Eq.~(\ref{gpe_labels}) is akin to performing the Madelung transformation.  

We now perform the same relabeling transformation that gives the circulation theorem and helicity conservation in Euler fluids, to seek analogous conservation laws. It is easily verified that the relabeling transformation: $a^i\rightarrow \tilde{a}^i = a^i + \epsilon\,\eta^i$, such that ${\partial\eta^i}/{\partial\tau} = 0, {\partial\eta^i}/{\partial a^i}=0$, is a symmetry of the above action $\tilde{S}_{\textrm{gpe}}$. The corresponding Noether charge is found to vanish identically, independent of $\eta^i$, as shown below:
\begin{align}
\mathcal{Q}_{gpe} &= \int d^3a \,\eta^j\left( \frac{\partial x^i}{\partial\tau}\frac{\partial x^i}{\partial a^j} - \frac{\partial \phi}{\partial a^j}\right) \nonumber \\
&= \int d^3a \,\eta^j\left( u^i\frac{\partial x^i}{\partial a^j} - \frac{\partial \phi}{\partial a^j}\right) \nonumber \\
&= \int d^3a \,\eta^j\left( \frac{\partial \phi}{\partial x^i}\frac{\partial x^i}{\partial a^j} - \frac{\partial \phi}{\partial a^j}\right)  \qquad \left(\because \mathbf{u}=\nabla\phi \right) \nonumber \\
&= \int d^3a \,\eta^j\left( \frac{\partial \phi}{\partial a^j} - \frac{\partial \phi}{\partial a^j}\right) = 0 \label{noether_gpe}
\end{align}

The above calculation suggests that the conserved charges analogous to helicity, and circulation trivially vanish for superfluids.

Note that the presence of an additional phase term $(-\partial_\tau\phi)$ in addition to the terms present in the Euler action $S_\textrm{Euler}$, is necessary to ensure Galilean invariance (as defined in \cite{sulem_nonlinear_2004}) of the modified action $\tilde{S}_\textrm{gpe}$, much like the constant term $(-c^2)$ \cite{kambe_elementary_2007} is necessary to ensure Galilean invariance of the classical fluid action. The presence of the additional phase term gives rise to mass conservation in the original Gross-Pitaevskii action $S_\textrm{gpe}$, which is manifestly Galilean invariant. However, mass conservation is inherent to the description of the superfluid when expressed in terms of the particle label co-ordinate frame $(\mathbf{a},\tau)$, and instead this term now has the effect of giving a vanishing conserved charge corresponding to relabeling symmetry transformations. We note that an alternative calculation due to Bretherton \cite{bretherton_note_1970} which derives the conservation of circulation using Hamilton's principle, also yields a vanishing conserved quantity in superfluids. 

\end{document}